\documentclass[fleqn,usenatbib]{mnras}

\usepackage[T1]{fontenc}
\DeclareRobustCommand{\VAN}[3]{#2}
\let\VANthebibliography\thebibliography
\def\thebibliography{\DeclareRobustCommand{\VAN}[3]{##3}\VANthebibliography}

\usepackage{graphicx}	
\usepackage{amsmath}	
\usepackage{amssymb}	
\usepackage{anyfontsize}
\usepackage{newtxtext,newtxmath}
\usepackage{mathtools}

\newcommand{\source}{{MAXI~J1820+070}}
\newcommand{\nicer}{NICER}
\newcommand{\hxmt}{\textit{Insight}-HXMT}

\title[Broad-band spectral-timing]{Broad-band spectral-timing: simultaneous NICER and HXMT observations reveal an anticorrelation between the softest and hardest X-ray fluxes in MAXI J1820+070}

\author[Bollemeijer et al.]{
Niek Bollemeijer,$^{1,2}$\thanks{E-mail: niekbollemeijer@proton.me}
Phil Uttley$^{1}$ and
Bei You$^{3}$
\\
$^{1}$Anton Pannekoek Institute for Astronomy, Amsterdam, Science Park 904, NL-1098 NH, The Netherlands\\
$^{2}$Dr. Karl Remeis-Observatory and Erlangen Centre for Astroparticle Physics, Sternwartstr. 7, 96049 Bamberg, Germany\\
$^{3}$School of Physics and Technology, Wuhan University, Wuhan 430072, People’s Republic of China\\
}

\date{Accepted XXX. Received YYY; in original form ZZZ}

\pubyear{2024}

\begin{document}
\label{firstpage}
\pagerange{\pageref{firstpage}--\pageref{lastpage}}
\maketitle

\begin{abstract}
Black hole X-ray binaries (BHXRBs) in the hard state show variable X-ray emission across a wide range of energies. Hard X-rays are thought to be produced through Compton-upscattering of seed photons from the accretion disc by the hot X-ray corona, leading to a spectral component roughly shaped like a power-law with a high-energy cut-off. A complete study of the coronal variability requires a full view of the spectrum, from the low-energy cut-off at the seed photon energy to the high energy cut-off, which may be related to the coronal temperature. We present a detailed spectral-timing study of strictly simultaneous observations by \nicer{} and \hxmt{} of the bright BHXRB MAXI~J1820+070, covering the 0.3-250 keV energy range. The high data quality allows us to study the lags and coherence across the full X-ray band. We find that at low Fourier frequencies ($\lesssim0.02$ Hz), the soft X-ray flux below 1 keV and hard X-ray flux above 100 keV are anticorrelated, i.e. phase lags of $|\pi|$ rad. No anticorrelation is observed when comparing energy bands above 1 keV. We consider several explanations for the observed complex behaviour, such as a modulation of the coronal cut-off energy by the disc, power-law pivoting and a contribution from non-thermal electrons. Future, more detailed modelling efforts based on the observed effects may place new constraints on the dynamics of the X-ray emitting regions around accreting black holes.
\end{abstract}




\section{Introduction}

The X-ray emission from black hole X-ray binaries (BHXRBs) in the hard state spans a wide range of energies. In the hard state of a BHXRB in outburst, the X-ray spectrum is dominated by a spectral component that is produced by Compton-upscattering of lower-energy seed photons \citep{Wardzinski_2002,Zdziarski_2004,Done_2007}. The Comptonized emission has the spectral shape of a power-law with a cut-off at $\sim100$ keV \citep{you2023apj}. The power-law emission originates in the corona, which is the name for a cloud of Compton-upscattering, hot electrons. The coronal geometry is subject to debate, with models ranging from a hot inner flow (e.g. \citealt{Ferreira_2006,Veledina_2016,Kawamura_2022,You_2023}) to the base of the radio jet (e.g. \citealt{Markoff_2005,Kylafis_2008}). The high-energy cut-off is associated with the coronal temperature \citep{Sunyaev_1979}, although it could also be related to the bulk velocity of electrons in the corona, if they are heated through magnetic reconnection \citep{Beloborodov_2017,Sironi_2020}. Beyond the high-energy cut-off, a power-law tail has been observed in the spectrum, which is attributed to a non-thermal population of electrons \citep{McConnell_2002,Poutanen_2009,Cangemi_2021,Zdziarski_2021}.

The main source of seed photons in the brighter states of an outburst is thought to be the soft X-ray emission from the accretion disc, which has a multi-colour blackbody spectrum with a temperature of $\lesssim0.5$ keV in the hard state \citep{Done_2007}. At lower luminosities, synchrotron radiation from the strongly magnetized corona may also be an important source of seed photons \citep{Sobolewska_2011,Veledina_2013hotflow,Skipper_2016,Yang_2025}. Observed spectra also show features related to the reflection of coronal photons on the disc, such as the iron line and Compton hump \citep{George_1991,Garcia_2014,Dauser_2016,You_2021}. Together, the disc, Comptonization and reflection components can be used to describe the energy spectrum reasonably well (see e.g. Fig. \ref{fig:spec_example}), but modelling the spectrum alone leaves degeneracies concerning the nature and geometry of the corona.

During the hard state, the X-ray flux is highly variable and can be studied using Fourier transforms. Large amplitude ($\sim5-20 \%$ rms) quasi-periodic oscillations (QPOs) are often observed (see \citet{Motta_2016} and \citet{Ingram_2019review} for recent reviews). Aperiodic variability known as broad-band noise is also observed, spanning a wide range of time-scales. Broad-band noise is often modelled with accretion rate fluctuations propagating through the accretion flow \citep{Lyubarskii_1997,Kotov_2001,zhan2025}. The observed variability over a broad range of time-scales may be produced if the fluctuations are produced on the viscous time-scale of each radius and propagate inward towards the black hole. The inner regions of the accretion flow emit harder X-rays, leading to the variability being observed over a wide range of energies.
In this framework, the disc both provides seed photons for the corona and is an important source of the variability observed in the coronal flux, e.g. via variations in seed photon flux and coronal heating \citep{Uttley_2025}. To improve our understanding of how the disc and corona interact, it is necessary to analyse the full X-ray spectrum, including the disc and the coronal emission up to the high-energy cut-off.

Studying the correlated variability at different X-ray energies is known as spectral-timing. Two main spectral-timing properties are the time-scale-dependent lags and the coherence between energy bands \citep{Uttley_2014review}. In the context of broad-band noise variability, the hard lags between bands with and without significant disc emission on time-scales of seconds have been attributed to accretion rate fluctuations, which are first seen in the disc and later in the corona \citep{Arevalo_2006,Ingram_2011,you2025}. Comparing energy bands within the coronal power-law range, hard lags are measured over a broad range of frequencies. These hard lags have been explained by Comptonization delays within an extended corona \citep{Kazanas_1997,Bellavita_2022}, but they may also be due to spectral pivoting of the power-law \citep{Kotov_2001,Uttley_2025} or propagation through a stratified corona \citep{Kawamura_2022}. In many sources, soft lags between soft X-rays below 1 keV, where the accretion disc contributes significantly to the spectrum, and harder bands are measured on short time-scales of $\sim0.1$ s \citep{Uttley_2011,DeMarco2017,Wang_2022_revmachine}. Such lags may originate from coronal power-law emission being emitted towards the disc, which will be reprocessed and lead to a reflection signature in the spectrum (e.g. \citealt{zhan2025}). The reflected emission has travelled a longer path to reach the observer and will therefore lag behind the power-law continuum, causing soft `reverberation' lags on the order of the average light-travel time between the disc and the corona, probing the coronal size and geometry \citep{Uttley_2014review,Ingram_2019reltrans,Kara_2019,De_Marco_2021,Mastroserio_2021}. The first detection of the Compton hump reverberation feature from MAXI~J1820+070 was reported by measuring lags in the broad energy range of $\sim$ 1-150 keV \citep{you2025}.

Alternatively, in a recent model \citep{Uttley_2025}, variations in seed photon and heating luminosity due to propagating fluctuations in the disc lead to pivoting of the coronal spectrum and associated hard lags. In this model, large soft lags on short-timescales naturally arise due to the delay between disc seed photons which drive soft power-law variability, and the disc reverberation signal which is produced when propagating fluctuations reach and heat the corona. 

The cross-spectral coherence is less straightforward to interpret. The coherence is defined between 0 and 1 and describes the degree of linear correlation between variability in different energy bands \citep{Vaughan_1997coherence,Nowak_1999}. Not only does the coherence determine the errors on the lags \citep{Bendat_2000}, but the coherence itself may also tell us about physical processes taking place in the accretion flow, although it is generally not modelled explicitly. Past studies have revealed that the coherence is often low when including (very) soft X-ray energies (notably in bands containing significant disc emission) (e.g. \citealt{Koenig_2024,Bollemeijer_2025}). Recently, \citet{Yang_2025} measured the coherence and covariance for high energies up to $\sim$ 150 keV, beyond the cut-off energy of the coronal power-law. A clear drop in coherence above 30 keV on both short- and long-timescales was found, relative to the
2–10 keV reference band. For non-unity coherence, lags are more difficult to interpret, as combinations of different mechanisms may lead to similar measured lags. 

None of the models mentioned earlier take into account non-linearly correlated variability (leading to reduced coherence), which we may expect to be present at the hardest X-ray energies, around the power-law cut-off. For example, if variations in seed photon and heating luminosities lead to pivoting of the power-law (as explained by \citealt{Uttley_2025}), we may expect the cut-off energy to be modulated as well, although no detailed modelling of such an effect has been done.

To explore the possible connections between the disc and coronal emission up to and above the cut-off energy, we here investigate the correlated X-ray variability across the full X-ray band. To gain insight into variability across a broad range of energies, we use strictly simultaneous observations by the \textit{Neutron Star Interior Composition ExploreR} (\nicer) and the \textit{Hard X-ray Modulation Telescope} (\hxmt) telescopes of the bright BHXRB \source, providing a detailed spectral-timing view from 0.3-250 keV for the first time.

\source{} went into outburst in March 2018 \citep{Nakahira_2018} and stayed in the hard state for several months. Its brightness (> 1 Crab for a significant part of the outburst) and low interstellar absorption ($\sim1.3\times10^{21}\rm{cm}^{-2}$, \citealt{Wang_2021,Lucchini_2023}) enables a clear view of the soft X-ray emission from the accretion disc below 1 keV. \source{} was monitored closely by both \nicer{} and \hxmt. Although no dedicated simultaneous observations were carried out, the close coverage of the source by both telescopes has led to a total of over 30 ks of simultaneous observations. Strictly simultaneous observations are required when doing Fourier-based timing analysis.

In this work, we first introduce the data and show how we obtained strictly simultaneous observations of \source{} with \nicer{} and \hxmt{} in Section \ref{sec:obs_methods}. In Section \ref{sec:lag_coh_freq}, we show time-scale dependent spectral-timing properties for different energy-bands. We focus on the energy-dependence of the low-frequency ($\lesssim$0.02 Hz) lags and coherence in Section \ref{subsec:lag_coh_spec}, revealing an anticorrelation between the softest and hardest X-rays on relatively long timescales of $\gtrsim50$ s. In Section \ref{sec:discussion}, we discuss several potential explanations for the observed anticorrelation and the complex coherence pattern. We consider power-law pivoting, coronal temperature changes and the presence of an extra spectral component at high energies, although more detailed modelling is required to reach more solid conclusions on the origin of the observed properties.

\section{Observations and methods}
\label{sec:obs_methods}
\subsection{Data}
We present a spectral-timing study of simultaneous \nicer{} and \hxmt{} observations of \source{} in the hard state of its outburst in March 2018. The simultaneous observations are listed in Table \ref{table:J1820data}. To find these data sets of simultaneous observations, we downloaded and reprocessed all hard state observations of \source{} by \nicer{} and \hxmt{} obtained between March 12 and July 5, 2018. Most of the simultaneous observations were made in the early stages of the outburst and we focus on the data in epoch 2 for the rest of the paper.

\nicer{} ({Neutron Star Interior Composition ExploreR}) is an X-ray telescope operated by NASA and located on the Internation Space Station (ISS) and was launched in 2017 \citep{Gendreau_2016}. \nicer{} consists of 52 Focal Plane Modules (FPMs), providing a large effective area, while its small deadtime and high throughput make it a flagship for X-ray spectral-timing studies. \nicer's soft response (0.2-12 keV) has highlighted the importance of including energy bands with a significant contribution from the accretion disc when studying the variability in BHXRBs \citep{Wang_2022_revmachine,Bollemeijer_2024,Koenig_2024}, which was first observed in XMM-Newton observations of fainter sources \citep{Uttley_2011,Cassatella_2012}.

We reprocessed the \nicer{} data using  the \texttt{nicerl2} command in HEASoft v6.35.1 \citep{FTOOLS_2014}, with default settings. \source{} was a very bright source, and its low interstellar absorption combined with \nicer's soft response gave rise to high count rates, causing telemetry saturation in \nicer. Some FPMs are affected more than others, and those are excluded by \texttt{nicerl2}. We corrected the light curves for the number of FPMs that were included. We describe our approach to mitigate further effects of telemetry saturation in Appendix \ref{app:gaps}. After reprocessing, we barycentered the cleaned event files using \texttt{barycorr}, with ephemeris DE440 \citep{Park_2021}. We set \texttt{barytime=NO}, which means that \texttt{barycorr} also automatically barycenters the Good-Time-Intervals (GTIs). We corrected the source coordinates such that they are exactly the same for both instruments when barycentering, since a mismatch in source coordinates can lead to significant time offsets between the instruments. 

The Hard X-ray Modulation Telescope (\hxmt) is China's first X-ray observatory and was also launched in 2017 \citep{Zhang_2014}. It consists of three instruments, covering the low energy (LE, 2-10 keV), medium energy (ME, 10-35 keV) and high energy (HE, 28-250 keV) ranges. The large effective area of the HE instrument in particular provides a unique view of hard X-rays in accreting black holes, especially for bright sources such as \source, which was observed extensively by \hxmt{} (e.g. \citealt{You_2021,Fan_2024}).

We applied the \texttt{hpipeline} functionality from HXMT Data Analysis Software (HXMTDAS) v2.06 to all \hxmt{} observations of \source. After reprocessing, we applied the \texttt{hxbary} command to the screened event files for the ME and HE instruments, also with ephemeris DE440. Because \texttt{hxbary} does not correct the GTIs in the event files, we shifted the GTIs by the difference between the barycentered and unbarycentered arrival time of the photon arriving closest to each GTI edge.

We compared the barycentered GTIs for all \nicer{} and \hxmt{} data and report the observations with at least 1 ks of simultaneous data in Table \ref{table:J1820data}. It is clear from Table \ref{table:J1820data} that most of the simultaneous data is concentrated in the second week of the outburst. Because the spectral hardness is also high at that point in the outburst, leading to high-quality hard X-ray data, we focus on data simultaneous with \nicer{} observations 1200120107-1200120112, corresponding to \hxmt{} observations P0114661003-P0114661006.

During the overlapping parts of GTIs, we created light curves using the clean event files for \nicer{} and the screened event files for ME and HE on \hxmt. In \nicer{} data, we ignore the background, as it is negligible (on the order of $\sim$ 1 count $\rm{s}^{-1}$, \citealt{Remillard_2022}) for sources with a count rate on the order of $10^4$ count $\rm{s}^{-1}$. For \hxmt, we created background light curves for the different energy bands. Because the background light curves produced with \texttt{hpipeline} are not barycentered, we shifted the background light curves by an interpolation of the shift in
event arrival times between barycentered and unbarycentered data at the nearest GTI start and stop times. Doing so returns a good estimate for the background, which is only modelled on timescales of tens of seconds in both the ME and HE instrument \citep{Guo_2020,Liao_2020HE}.

\begin{figure}
    \centering
    \includegraphics[width=\linewidth]{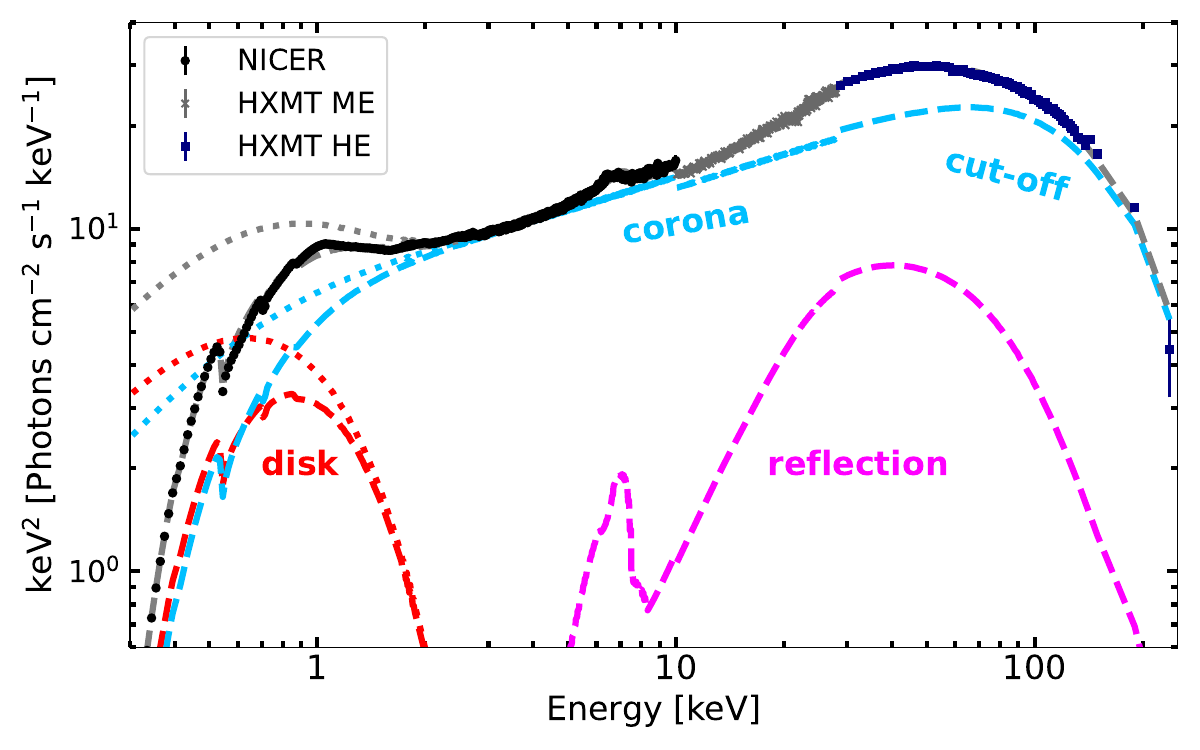}
    \caption{Example energy spectrum of a subset of data from NICER ObsID 107 and HXMT observations 0301 with a simple fit for illustrative purposes. The shaded areas correspond to the energy bands used in the lag vs frequency spectra in Fig. \ref{fig:stprops_0.3-0.5_3-10keV}. The dashed and dotted lines show the absorbed and unabsorbed model components, respectively, and the full model is shown in grey.}
    \label{fig:spec_example}
\end{figure}

For clarity, we show a broad-band X-ray energy spectrum with \nicer{} and \hxmt{} data in Fig. \ref{fig:spec_example}. The \nicer{} spectrum was extracted using \texttt{nicerl3-spect} with default settings, while \texttt{hpipeline} provided the \hxmt{} spectral files. We only show quasi-simultaneous spectra of \nicer{} ObsID 107 and HXMT observation 0301 in Fig. \ref{fig:spec_example}.

\subsection{Methods}

We use the simultaneous \nicer{} and \hxmt{} observations to study the energy-dependence of three spectral-timing properties over a very broad range of energies. We focus on the power spectra in various energy bands, and the phase lags and coherence between different energy bands.

For the power spectra, we use the fractional rms normalization \citep{Belloni_1990,Miyamoto_1992}. When comparing the fractional rms variability in different energy bands, the background contribution can artificially lower the measured rms. The background is particularly important at energies > 150 keV in HE of \hxmt. At low frequencies, around 0.01 Hz, corresponding to timescales of around 100 seconds, the background itself may be variable and contribute power that is not due to the source, which is why we subtract background light curves from the data and show power spectra for background-subtracted light curves.

Phase lags are defined as the phase of the cross-spectrum averaged over many light curve segments and we follow \citet{Uttley_2014review} to obtain them. When calculating phase lags in different narrow energy bands, it is common to choose a broad reference band to maximize the signal-to-noise ratio \citep{Ingram_2019error}. We show that for the very high data quality of observations of \source, using narrower reference bands yields good measurements of the phase lags as well, and more importantly, we show that the choice of reference band has large effects on the measured phase lags, especially the very soft and hard regions of the X-ray spectrum. The errors on the frequency-dependent lags are calculated using the raw coherence \citep{Bendat_2000,Uttley_2014review}. For the errors in lag-energy spectra, we use the prescription in \citet{Ingram_2019error}, which take into account that different energy bands contain correlated variability, but also discuss why they can lead to overestimation of the errors in some cases in Appendix \ref{sec:app_coherr}. 

\begin{figure*}
    \centering
    \includegraphics[width=\textwidth]{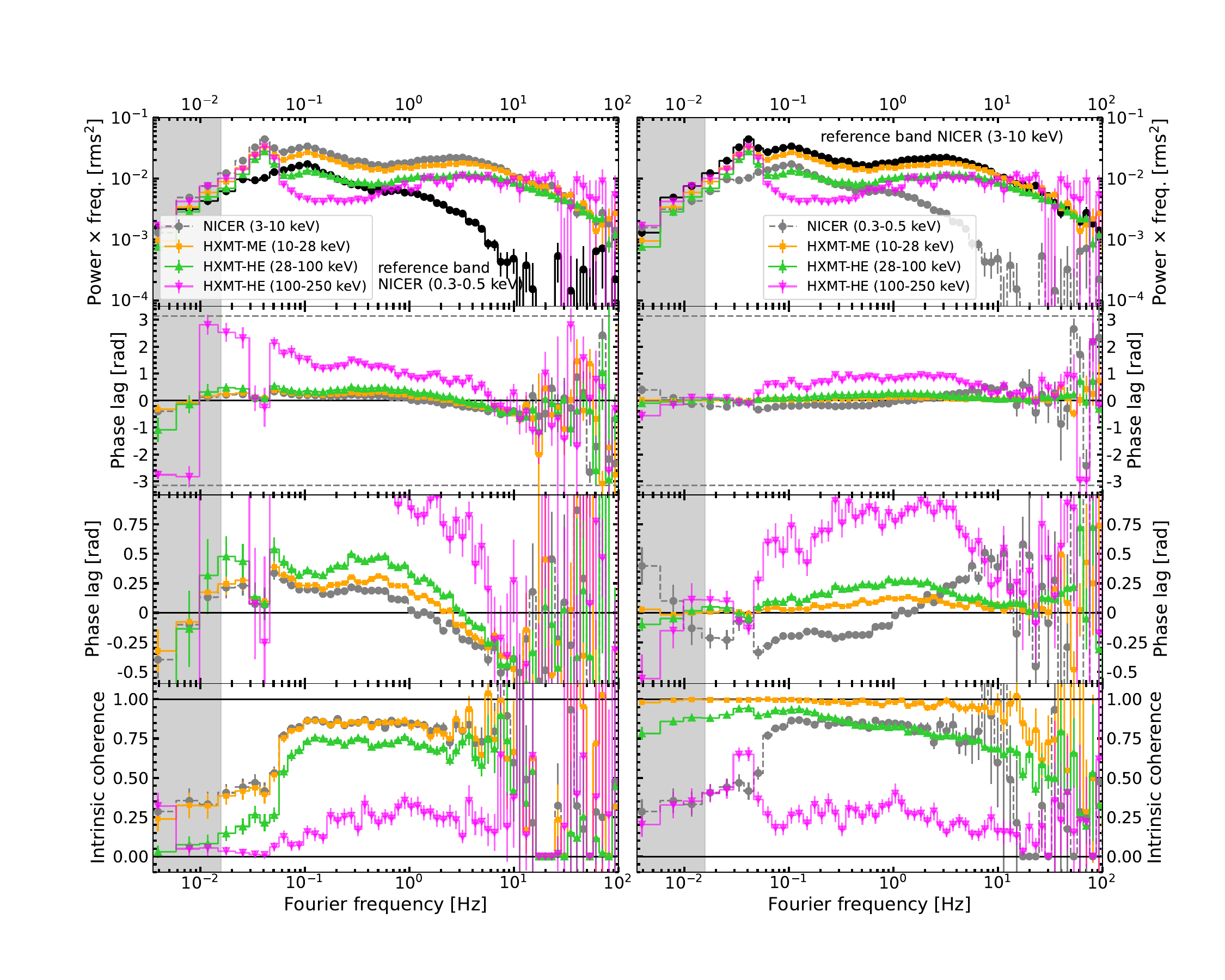}
    \caption{The upper panels show power spectra with 5 different energy bands using simultaneous data from \nicer{} and \hxmt, with \nicer{} ObsIDs 1200120107-1200120112 and \hxmt{} observations P011466100301-P01146610601, made between 22 and 27 March 2018 (epoch 2, see also table \ref{table:J1820data}). The solid black power spectra correspond to the reference bands (0.3-0.5 keV for the left column, 3-10 keV for the right column) for the phase lags and coherence shown in the middle and lower panels. The phase lags are shown in two panels for both columns, with the upper-middle panel y-axis including $-\pi$ to $\pi$ rad (dashed grey lines,) while the lower-middle panel zooms in on the lags between -0.6 to 1 rad. For the phase lags in the middle right panels, the sign of the lags between the 0.3-0.5 and 3-10 keV bands (dashed grey) are inverted with respect to the conventional way of plotting them, to keep the sign of lags consistent with the other energy bands: negative lags are hard lags for the 0.3-0.5 keV subject band in this particular case, while all other bands follow the convention that hard lags are positive. The power spectra below the QPO frequency look similar in all energy bands, but for the 0.3-0.5 keV reference band, the coherence is low and the phase lags indicate an anticorrelation with the highest energy band of 100-250 keV. When using the harder 3-10 keV reference band shown in the right column, all harder subject bands are much more coherent than for the softer reference band and the anticorrelation at low frequencies is not visible. For all combinations of energy bands, the lags are small at the QPO frequency. The Poisson noise-dominated high-frequency coherence is not well defined, particularly for the soft reference band, and is set to zero when noise subtraction causes the calculated coherence to be negative. The grey shaded areas correspond to the frequency range used in Figs. \ref{fig:lagcoh1} and \ref{fig:colormeshlog}.}
    \label{fig:stprops_0.3-0.5_3-10keV}
\end{figure*}

The coherence measures the fraction of the variance in two light curves that can be obtained from a linear transformation between the two time series \citep{Vaughan_1997coherence}. Non-unity coherence generally arises when the variability consists of multiple independent components, which have different contributions in both energy bands. This includes the extreme case where a variability component is present in one energy band and not the other. The coherence can also be reduced due to non-linear correlations between different energies \citep{Nowak_1999}. The raw coherence is the value obtained when not accounting for Poisson noise, which (in the absence of shared deadtime) is independent in different detectors. By subtracting the contribution from Poisson noise from power spectra, we can estimate the intrinsic coherence. We calculate the errors on the coherence following \citet{Vaughan_1997coherence}, although we note that the prescriptions are not 
valid for low intrinsic coherence and poor signal-to-noise, as is further discussed in Appendix \ref{sec:app_coherr}.

In this paper, we use 256 s segments at 1/512 s time resolution, yielding a Fourier frequency range of 1/256 to 256 Hz. For the high source count rates, deadtime can be significant and lower the Poisson noise level, especially in the ME instrument \citep{Cao_2020}. We therefore estimate the Poisson noise level by fitting a constant to the power spectrum above 200 Hz, where the source variability is negligible.

\section{Results}
\label{sec:results}
\subsection{Lag- and coherence versus frequency}
\label{sec:lag_coh_freq}

First, we show in Fig. \ref{fig:spec_example} an example of a full X-ray spectrum from the observations used, which includes contributions from the accretion disc and at the high-energy cut-off. The components of a fit with a simple, descriptive model (\texttt{tbabs*(diskbb+nthcomp+relxillCp)*constant}) are also shown. The \texttt{tbabs} component accounts for interstellar absorption \citep{Wilms_2000}, \texttt{diskbb} describes thermal emission from the accretion disc \citep{Mitsuda_1984}, \texttt{nthcomp} the Comptonized emission from the corona \citep{Zdziarski_1996,Zycki_1999}, \texttt{relxillCp} the reflection spectrum \citep{Garcia_2014,Dauser_2014} and the \texttt{constant} component takes into account small normalization differences between the instruments used (visible as slight discontinuities between ME and the other instruments). A detailed spectroscopic analysis lies outside the scope of this work and the spectral components are only shown to illustrate the spectral decomposition we assume to interpret the timing results. We therefore do not show residuals or discuss the fitted parameters. From Fig. \ref{fig:spec_example}, it is clear that the data used cover the full X-ray spectrum, from the thermal disc emission to the high-energy cut-off.

The coloured bands correspond to the energy bands used in Fig. \ref{fig:stprops_0.3-0.5_3-10keV}, where we show power spectra, phase lag and coherence versus frequency for different energy bands. The data shown are obtained from the simultaneous parts of ObsIDs 107 to 112 of \nicer, made between 22 and 27 March 2018 (epoch 2 in table \ref{table:J1820data}). For epoch 2, the coronal electron temperature is about 60 keV, when measured with a simple model based on \texttt{relxillCp} with \hxmt{} data (see Fig. \ref{fig:spec_example} and \citealt{You_2021}). The difference between the columns in Fig. \ref{fig:stprops_0.3-0.5_3-10keV} is the choice of reference band for the phase lags and coherence functions, which is very soft (0.3-0.5 keV) for the left column of Fig. \ref{fig:stprops_0.3-0.5_3-10keV} and harder (3-10 keV) for the right column. Because there is a small shift in the QPO frequency from $\sim0.037$ to $\sim0.05$ Hz during the observation days that are combined here, the QPO peak is broadened, but still clearly visible in all energy bands except the softest band. As was reported by \citet{Ma_2021}, the QPO rms is similar for all energies above 1 keV. The 0.3-0.5 keV band does not show a peak at the QPO fundamental frequency, but it does show excess variability at the second harmonic frequency around 0.1 Hz. Due to the drift in QPO frequency, however, the harmonic is not clear in the power spectra shown here. The power in the softest band is suppressed at frequencies above a few Hz, which may be due to viscous diffusion \citep{Churazov_2001,Mushtukov_2018} and/or due to the contribution to the total soft band emission from larger disc radii, which should show less variability at high frequencies \citep{Uttley_2025}. The hardest band, 100-250 keV, has a clear QPO fundamental, but no obvious harmonic signal and the variability is suppressed in the 0.1 to 10 Hz range compared to softer power-law bands. 

When we compare the phase lag versus frequency spectra for different energy bands (middle panels), the following aspects stand out. For most combinations of energy bands, the lags have canonical properties: hard lags between harder bands above a few keV, where the energy spectrum is dominated by the coronal power-law, especially between 0.1 and 10 Hz (right column of Fig. \ref{fig:stprops_0.3-0.5_3-10keV}) and soft lags between soft bands (below 1 keV, possibly related to disc emission) and harder bands at high frequencies above a few Hz (see left column of Fig. \ref{fig:stprops_0.3-0.5_3-10keV}), as was observed for \source{} by \citet{Kara_2019}, \citet{De_Marco_2021}, \citet{Wang_2021} and \citet{Bollemeijer_2024} using \nicer{} data only. The large phase lags above the QPO frequency for the hardest energies >100 keV were also observed before in \hxmt{} observations of \source, e.g. by \citet{Ma_2021}, \citet{Kawamura_2023} and \citet{you2025}. 

The phase lags at the QPO fundamental frequency itself are small and often consistent with zero for all combinations of energy bands, as was also found to be the case in Swift~J1727.8-1613 for a broad range of QPO frequencies \citep{Bollemeijer_2025}. Below the QPO frequency, however, we see a large dependence of the phase lags on the reference band chosen. For the right column of Fig. \ref{fig:stprops_0.3-0.5_3-10keV}, where we use a 3-10 keV reference band, the lags are close to zero. For the soft 0.3-0.5 keV reference band (left column), we see that the phase lags are very large and show phase wrapping around $|\pi|$ rad. These very large lags indicate the presence of an anticorrelation between the softest X-rays (possibly from the disc) and the hardest X-rays above the coronal cut-off energy. For these observations, the measured inner disc temperature is about 0.25 keV \citep{Fan_2024}, while the electron temperature was determined to be around 60 keV based on fits with \texttt{relxillCp} (see \citet{You_2021} and Fig. \ref{fig:spec_example}). In the reflection model \texttt{relxillCp}, the coronal continuum emission is approximated with the model \texttt{nthcomp}, which has an approximately exponential rollover at about 2-3 times $kT_e$. The lags of $|\pi|$ rad are the main finding investigated in this work, as they indicate an anticorrelation between different energy bands. 

\begin{figure}
    \centering
    \includegraphics[width=\linewidth]{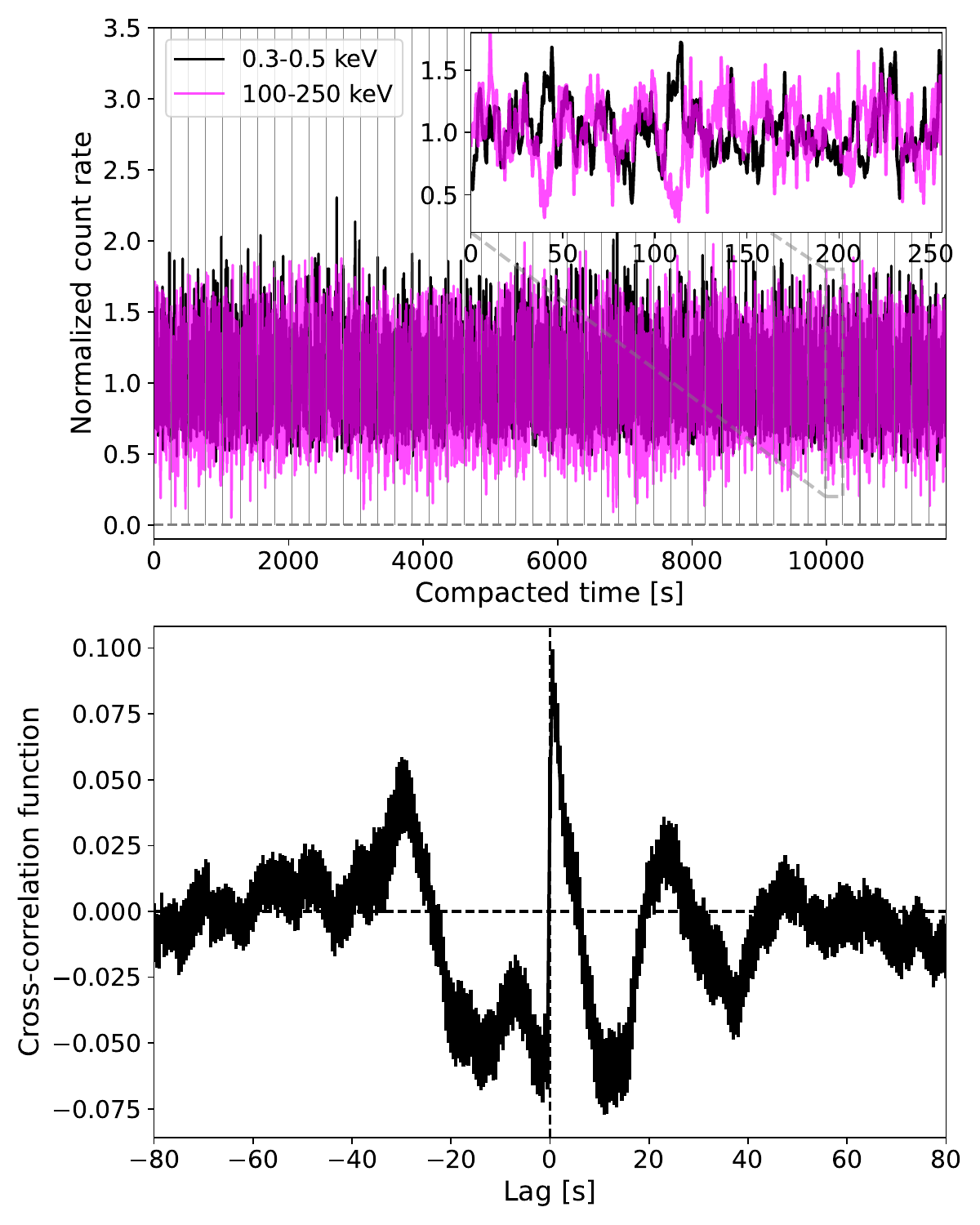}
    \caption{The upper panel shows the normalized and compacted light curve of the 0.3-0.5 and 100-250 keV bands in epoch 2, rebinned to a time resolution of 0.5 s. The vertical lines show the edges of each segment and the average flux does not change significantly. In the inset plot, it is clear that the fluxes of these two energy bands are anticorrelated during a few flares. This is also visible in the average cross-correlation function, which shows a broad (roughly -20 to 20 s) negative trough centred around 0 s lag. The negative CCF switching to a sharp positive peak at 0 s lag corresponds to the hard lags on shorter time-scales ($\sim$0.1-10 Hz in Fig. \ref{fig:stprops_0.3-0.5_3-10keV}). 1-$\sigma$ error bars obtained by bootstrapping are shown for the CCF.}
    \label{fig:lc_ccf}
\end{figure}

More evidence for the presence of an anticorrelation is found by inspecting the light curves themselves, as is shown in the upper panel of Fig. \ref{fig:lc_ccf} for the 0.3-0.5 and 100-250 keV energy bands of epoch 2. The upper panel shows that the soft band flux often peaks when the hard band shows a minimum. The inset plot of a single 256 s segment contains several soft flares accompanied by dips in hard flux. The lower panel in Fig. \ref{fig:lc_ccf} contains the average cross-correlation function (CCF) of the two light curves for epoch two. Negative values of the CCF indicate anticorrelated behaviour and the broad negative structure between -20 and 20 s lags confirms that on time-scales of tens of seconds, the softest and hardest bands are anticorrelated. The positive peak in the CCF at 0 s lag corresponds to the the large hard lags on shorter timescales ($\sim$ seconds) also visible in the lag vs frequency spectra in Fig. \ref{fig:stprops_0.3-0.5_3-10keV}.

The lower panels of Fig. \ref{fig:stprops_0.3-0.5_3-10keV} show the coherence for the different energy bands. As we showed in \citet{Bollemeijer_2025varlags} with \nicer{} data only, the coherence between a very soft band (0.3-0.5 keV) and harder bands (3-10 keV) drops at and below the QPO frequency in \source. However, at very low frequencies, we observe a notable rise in the coherence between the hardest and softest bands, which may be related to the anticorrelation. Further interpretation of the anticorrelation requires higher spectral resolution of the spectral-timing properties than the broad energy bands used in Fig. \ref{fig:stprops_0.3-0.5_3-10keV}. In the next section, we therefore investigate the energy-dependence of the lags and coherence in more detail.

\subsection{Lag- and coherence spectra}
\label{subsec:lag_coh_spec}

\begin{figure*}
    \centering
    \includegraphics[width=\textwidth]{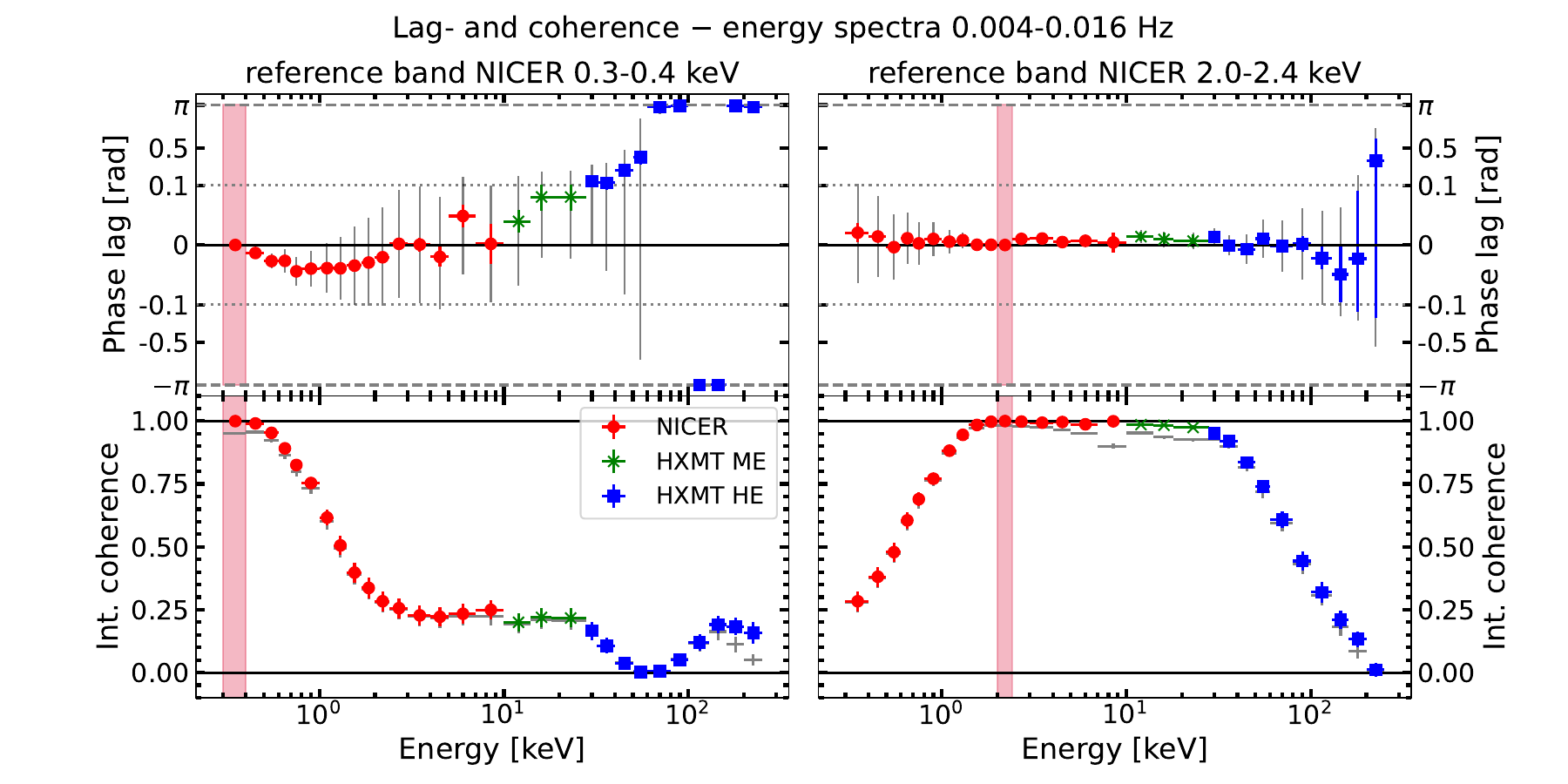}
    \caption{The lag- and coherence vs energy spectra of the data in epoch 2 for the $\frac{1}{256}$ ($\sim$0.004) to $\frac{1}{64}$ ($\sim$0.016) Hz frequency range for two different reference bands: in the left column the reference energy is 0.3-0.4 keV, while for the right column it is 2-2.4 keV, as illustrated by the shaded area. The phase lag y-axis is shown with a `symlog' scaling, which means that it is linear up to $|0.1|$ rad (indicated by the dotted grey lines) and log-scaled for larger absolute values. For the 0.3-0.4 keV reference band, the lags are close to 0 rad and the coherence decreases with energy up to $\sim30$ keV. Between 60 and 90 keV the lags flip to $\sim|\pi|$ rad and the coherence rises to $\sim0.2$, indicating an anticorrelation between the lowest and highest energies. The right panels, with reference band 2-2.4 keV, show lags that are consistent with 0 for all energies, while the coherence drops at low and high energies. The drop of the coherence at high energies was first reported by \citet{Yang_2025}. The phase lags have two different error bars, with the true errors having values in between those plotted. For an extended discussion on the errors, we refer to Appendix \ref{sec:app_coherr}. The grey points in the lower panels indicate the raw coherence.}
    \label{fig:lagcoh1}
\end{figure*}

\begin{figure*}
       \centering
    \includegraphics[width=\textwidth]{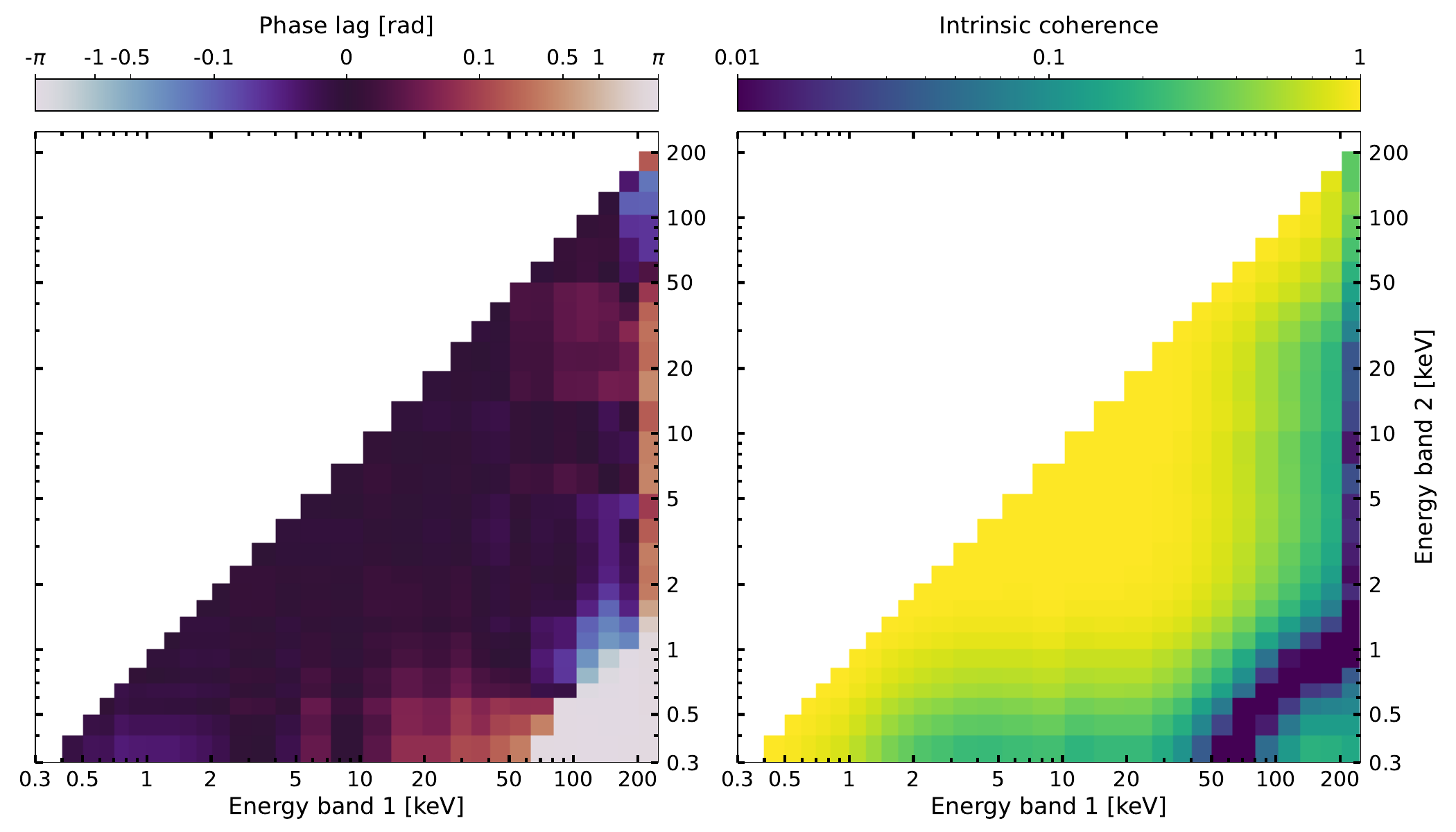}
    \caption{The panels show the lag and coherence versus energy spectra between $\frac{1}{256}$ ($\sim$0.004) and $\frac{1}{64}$ ($\sim$0.016) Hz for all combinations of energy bands, indicated on both axes. The phase lags (left panel) are shown on a `symlog' scaling, which is linear between -0.1 and 0.1 rad and logarithmic at larger absolute values. The colormap is cyclic and a phase lag of $\pi$ and $-\pi$ rad both correspond to an anticorrelation. The lags between any combination of energy bands are shown following the convention that positive lags always correspond to hard lags. The coherence (right panel) is shown on a logarithmic scaling, which makes the drop in coherence for very soft and hard bands (lower right corner) clearly visible. The coherence between all combinations of energy bands can be read in a similar way to the lags.
    For all reference bands below 1 keV, the hardest energy bands switch to a lag of $|\pi|$ rad. The anticorrelated energies depend on the soft band used, with an upward trend in the anticorrelated hard band for harder soft bands ($\lesssim$1 keV). The coherence shows a dip that follows the same trend for soft reference bands, which is also visible in Fig. \ref{fig:lagcoh1}. For reference bands above 1 keV, there is no anticorrelation and the coherence decreases when including energy bands below 1 keV or above $\sim$30 keV. The highest energy bands show large hard lags for all reference bands, although we note that the coherence is low, so the errors on the measurements are significant, as is also visible in the right panel of Fig. \ref{fig:lagcoh1}.}
    \label{fig:colormeshlog}
    \centering
\end{figure*}

In Fig. \ref{fig:lagcoh1}, we see that the choice of reference band strongly impacts the phase lags and coherence below the QPO frequency, especially for the hardest energy band, above 100 keV. To investigate the energy-dependence of the anticorrelation observed between very soft and very hard bands in the middle left panel of Fig. \ref{fig:stprops_0.3-0.5_3-10keV}, we present lag-energy spectra following \citet{Uttley_2014review} and \citet{Ingram_2019error}. We calculate phase lags and coherence in the 0.00390625 ($\frac{1}{256}$) to 0.015625 ($\frac{1}{64}$) Hz (referred to as 0.004 and 0.016 Hz from now on) range by averaging over the cross-spectrum and do so for a total of 29 energy bands: 16 for \nicer, 3 for ME and 10 for HE.\footnote{The energy bin edges are 0.3, 0.4, 0.5, 0.6, 0.7,  0.8, 1, 1.2, 1.4, 1.7, 2, 2.4, 3, 4, 5, 7 and 10 keV for \nicer, 10, 14, 18 and 28 keV for ME and 28, 32, 40, 50, 60, 80, 100, 130, 160, 200 and 250 keV for HE.} We created lag- and coherence versus energy spectra using each of the 29 energy bands as reference band and show two examples in Fig. \ref{fig:lagcoh1} (reference bands 0.3-0.4 and 2-2.4 keV with \nicer). 

In the upper left panel of Fig. \ref{fig:lagcoh1}, we see that the low-frequency phase lags with respect to the 0.3-0.4 keV band are small up to $\sim$40 keV, where we see a sharp rise. Above $\sim$70 keV, the phase lags are consistent with $|\pi|$ rad for all energy bands. The sizeable errors on the 0.3-0.5 vs 100-250 keV lags in Fig. \ref{fig:stprops_0.3-0.5_3-10keV} prevent us from directly drawing conclusions on the nature of the large lags. However, the fact that all energy bands >80 keV vs 0.3-0.5 keV show a phase lag of $|\pi|$ rad strongly suggests that there is an anticorrelation between the softest and hardest X-rays on the relatively long timescales we are probing here ($\sim50-250$ s), which is not simply due to phase-wrapping of a very large time lag. We show two different errors (thin black and thicker coloured error bars) on the phase lags of Fig. \ref{fig:lagcoh1}, while the true value must lie in between both. We discuss the errors in more detail in Appendix \ref{sec:app_coherr}.

The lower left panel of Fig. \ref{fig:lagcoh1} shows the low-frequency intrinsic coherence with respect to the 0.3-0.4 keV reference band. The raw coherence is shown in grey. There is a clear decrease in coherence as the subject energy band increases up to about 2 keV, where the coherence plateaus at a value of $\sim$ 0.25. Above 30 keV, the coherence drops further and is very low at around 70 keV, but partially recovers above 100 keV and reaches a value of $\sim$0.2 for the highest energies. The drop and recovery of the coherence at the same energies as where the phase lags flip from 0 to $\pi$ rad places constraints on the mechanism causing the observed behaviour, as we shall discuss in more detail in Section \ref{sec:discussion}.

When we shift the reference band to harder energies, e.g. 2-2.4 keV, as shown in the right column of Fig. \ref{fig:lagcoh1}, the lags and coherence show very different behaviour. The lags are consistent with zero across the full energy range. The coherence shows a clear decrease below 1 keV, again related to the drop in coherence seen below the QPO frequency in \citet{Bollemeijer_2025varlags}. In \citet{Yang_2025}, it was found that,
at higher energies and over a broad range of time-scales, the coherence with respect to the 2-10 keV energy band remains almost unity up to about 30 keV and then starts to drop, until it is close to zero for the hardest energy band (200-250 keV). The decrease in coherence at high energies is investigated in more detail by \citet{Yang_2025}.

The differences between the left and right panels of Fig. \ref{fig:lagcoh1} are so large that it is worth investigating how the low-frequency phase lags and coherence depend on the reference band energy in more detail, as shown in Fig. \ref{fig:colormeshlog}. Because Fig. \ref{fig:colormeshlog} contains a lot of information, we will explain in some detail how to read it. Both axes correspond to the energy bands used to calculate the lags (left panel) and coherence (right panel). The values of the phase lags and coherence are colour-coded, as shown by the colour bars. Except for the hardest and softest bands, the lags and coherence for any reference band can be read from the plot by following the plotted colours first in the vertical and then the horizontal direction. To give an example, to read the lags between a 2-2.4 keV reference (as also shown in the right panel of Fig. \ref{fig:lagcoh1}) and the other bands, we start on the x-axis at 2-2.4 keV and determine the lags for softer subject bands in the vertical direction until we reach the diagonal. The lags between the 2-2.4 keV reference band and harder energies are visible in the horizontal direction, where Energy band 2 (2-2.4 keV) is the reference band now. In Fig. \ref{fig:colormeshlog}, the plotted positive lags are always hard lags. The same approach can be used to determine the coherence.

 The phase lags have a `symlog' scaling, which is linear between -0.1 and 0.1 rad, to highlight small changes in the lags, while the white colors in the lower right corner indicate phase lags close to $|\pi|$ rad between very soft and very hard X-ray bands. The intrinsic coherence is shown on a log-scale down to 0.01. For reference bands below 1 keV, we clearly see the initial decrease in coherence for energies above 1 keV. The drop to very small values of coherence around 70 keV in the leftmost column of Fig. \ref{fig:colormeshlog} (reference band 0.3-0.4 keV and in lower left panel of Fig. \ref{fig:lagcoh1}) is clearly visible, as is the partial recovery of coherence at the highest subject band energies (>100 keV). It is also striking that, as the reference band energy increases, both the high-energy dip in coherence and the switch from $\sim0$ to $|\pi|$ rad phase lag shift linearly to harder bands as well. For reference bands above $\sim$1 keV, we observe a gradual decrease in coherence above $\sim$30-50 keV and the lags are small for almost all subject bands. The hardest subject band (200-250 keV) has large hard lags for reference bands $\gtrsim1.5$ keV, which are reminiscent of the lags that were observed by \citet{Basak_2025} using INTEGRAL data in \source. However, the low coherence (a few percent) for these combinations of energy bands leads to significant errors on the lags, complicating their interpretation. 

Finally, we mention that in Fig. \ref{fig:colormeshlog}, it is remarkable that in the coherence (above the diagonal), we can distinguish three energy ranges that are relatively coherent with each other and less coherent with other energies. The coherence is high between bands below 1 keV, between $\sim$1 and 80 keV and between 80 and 200 keV. We can also see that the X-ray fluxes below 1 keV and above 80 keV are partially coherent with each other and anticorrelated (lags of ~$|\pi|$ rad). We will discuss possible interpretations of the features observed in Fig. \ref{fig:colormeshlog} in the next section. 

\section{Discussion}
\label{sec:discussion}

In Section \ref{sec:results}, we showed various spectral-timing properties across a broad range of X-ray energies. Most of the phenomenology presented in Fig. \ref{fig:stprops_0.3-0.5_3-10keV} is canonical behaviour and has been observed before. We briefly summarise those properties. Hard lags between most combinations of energy bands across a wide range of frequencies, often attributed to propagating fluctuations in a hot flow \citep{Ingram_2011,Axelsson_2021,Kawamura_2023,zhan2025}. Between the softest reference band and harder bands, we observe soft lags at frequencies above a few Hz, often attributed to reverberation of coronal photons on the disc \citep{Fabian_2009,Uttley_2014review,Kara_2019,De_Marco_2021,Wang_2022_revmachine,zhan2025} or propagation effects \citep{Uttley_2025}. The lags at the QPO frequency are small and may be explained by precession of the corona leading to spectral pivoting \citep{Veledina_2013QPO, Ingram_2014, You_2018, You_2020} or Comptonization delays in an extended coroan \citep{Kazanas_1997,Bellavita_2022,MaRuican_2023}. 
\newline

We now summarize the findings that, to our knowledge, were not reported before.
\begin{enumerate}
    \item Between the very soft ($<$1 keV) and hardest bands, the phase lags are large over a wide range of frequencies, notably except at the QPO fundamental frequency. On relatively long time-scales of $\gtrsim$50 s (0.02 Hz), the phase lag between the very soft and hard (>100 keV) X-ray fluxes is $|\pi|$ rad, indicating an anticorrelation.
    \item When using a slightly harder reference band, above 1 keV, in which emission from the accretion disc is unimportant, we do not observe an anticorrelation on any time-scale.
    \item The hard energy band at which the anticorrelation becomes apparent increases with reference band energy.
    \item The switch from correlated to anticorrelated energy bands is accompanied by a drop in coherence, which partially recovers at harder energies.
\end{enumerate}

In the following subsections, we investigate scenarios that may lead to the observed lag and coherence behaviour, particularly the anticorrelation between very soft and hard X-rays. We note that the fact that the coherence is far from unity for many combinations of energy bands and Fourier frequencies. As a consequence, no single linear impulse response function can reproduce all data properties. The ideas presented in this section may therefore be read as a starting point for future, more complicated models.

We consider two different types of explanation. In the first, complex variability in a single spectral component leads to the anticorrelation, e.g. via changes in the cut-off energy or spectral pivoting. We argue that it is difficult to reproduce the observed behaviour with such scenarios in subsections \ref{subsec:T_changes} and \ref{subsec:pivoting}. The second explanation, outlined in subsection \ref{subsec:hybrid}, is based on the presence of independently varying spectral components, the relative strength of which determines the spectral-timing properties in each energy band. Although the (physical) details of such an explanation should be studied in more detail, doing so lies outside the scope of this work.

\subsection{Changes in coronal temperature}
\label{subsec:T_changes}

From Figs. \ref{fig:lagcoh1} and \ref{fig:colormeshlog}, it is clear that on time-scales of $\sim$50-250 s, flux variations at energies $\lesssim$1 keV are anticorrelated with those $\gtrsim$80 keV. In the coherence versus energy spectra, we only observe the anticorrelation at energies above bands that show a drop in coherence to very low values, which partially recovers, but remains far from unity. The fact that we only see the anticorrelation using very soft X-ray bands may indicate that it is related to emission from the accretion disc. Also, the anticorrelation is only observed on long time-scales, which is often linked with variations in accretion rate at relatively large radii in the disc, where the viscous time-scales are longer than closer to the black hole. In models based on accretion rate fluctuations, such variability is propagated to small radii and thus also visible at harder energies. 

If we assume that the corona emits the hard X-ray spectrum through thermal Comptonization, the coronal temperature may change due to external factors, such as cooling by increased seed photon flux from the variable disc. In this scenario, a rise in disc flux leads to cooling of the corona and thus a lower cut-off energy \citep{Uttley_2025}. If the cooling of the corona is sufficiently strong and leads to a large drop in the cut-off energy, the flux in the hardest energy bands (above the cut-off energy) would decrease. For an anticorrelation to be measurable, the decrease in flux at high energies has to be larger than the normalization increase expected from the higher seed photon flux. If that is the case, the result would be an anticorrelation between bands with significant disc emission and bands above the cut-off energy. 

However, this scenario is difficult to reconcile with the lack of an anticorrelation for reference bands slightly above 1 keV. An increase in seed photon luminosity is expected to lead to a rise in soft power-law emission from the corona \citep{Uttley_2025} and thus, the soft power-law emission should also be anticorrelated with the hardest energy bands. We only measure the anticorrelation between energies $\lesssim$1 keV and hard bands $\gtrsim$80 keV, while no anticorrelation can be observed for reference bands above $\sim$1.5 keV, which is inconsistent with the expected behaviour of the soft power-law emission. 
Also, the scenario where seed photons from the disc modulate the coronal cut-off energy does not naturally explain the linear relation between the energy of the anticorrelated hard X-rays and the reference band energy seen in Fig. \ref{fig:colormeshlog}. 

Possibly, the shift in contributions from the accretion disc and soft power-law to the total emission below 1 keV can explain that relation. At very high energies, the variability would arise from two competing effects. Part of the variability would be due to a varying coronal temperature, which is anticorrelated with disc emission and most important at the highest energies. The second type of variability is correlated with the soft power-law emission, e.g. due to energy-independent normalization changes in the coronal emission. These two types of variability are uncorrelated with each other, leading to the reduced coherence between the hardest bands and all softer X-ray bands.

As we move the reference band energy to higher values, the contribution from the soft power-law emission increases. For the variability component that is anticorrelated with the disc to dominate the measured lags, the subject band energy would also have to be higher. Such competition between different types of variability could lead to the linear relation between anticorrelated hard X-ray energies and the reference band energy observed.

\subsection{Power-law pivoting}
\label{subsec:pivoting}

Since changes in coronal cut-off energy due to seed photon variations are unlikely to cause the observed spectral-timing properties, we investigate other effects that may lead to an anticorrelation. From previous research, we know that the power-law emission exhibits pivoting (a varying photon index $\Gamma$) and both the disc and reflection parameters change on short time-scales (e.g. \citealt{Ingram_2016,Skipper_2016,Kawamura_2023}). 

Anticorrelations between different energy bands may be reproduced by a pivoting power-law. It has long been known that the power-law index changes as a function of luminosity in X-ray binaries (and AGN), with sources generally showing softer-when-brighter behaviour in luminous states \citep{Skipper_2016,you2023apj}. A pivoting power-law, parametrized by changes in both the photon index and the normalization, can lead to lags between power-law energies if both parameters do not vary simultaneously, which has been modelled extensively \citep{Mastroserio_2018, Ingram_2019reltrans}. A pivoting power-law can have a pivot point at the energy where variations in both parameters cancel each other, leading to a lack of correlated variability at that energy, while bands below and above the pivot point are anticorrelated. \citet{Uttley_2025} give an approximate expression for the pivot energy of a power-law that pivots due to variations in the ratio of seed photon to heating luminosities,

\begin{equation}
\label{eq:pivotenergy}
    \mathrm{ln} \left( 
    \frac{E_{\mathrm{piv}}}{E_{\mathrm{s}}} \right) =
    \left[ \beta \langle \Gamma \rangle \left( 
    1 - \frac{g_{\mathrm{h}}(\tau)}{\langle L_{\mathrm{h}} \rangle}/\frac{g_{\mathrm{s}}(\tau)}{\langle L_{\mathrm{s}}  \rangle}  
    \right)  \right]^{-1} + \frac{1}{\langle \Gamma \rangle -1},
\end{equation}

where $E_{\mathrm{piv}}$ and $E_{\mathrm{s}}$ are the pivoting and seed photon energies, $\beta=1/6$ (for BHXRBs, \citealt{Beloborodov_2001}), $\langle \Gamma \rangle$ is the time-averaged value of the photon index and  $g_{\rm{s}}(\tau)/\langle L_{\rm{s}}\rangle$ and  $g_{\rm{h}}(\tau)/\langle L_{\rm{h}}\rangle$ are the impulse responses of the seed photon and heating luminosities over their expectation values. Equation \ref{eq:pivotenergy} is based on the assumption that the power-law is not cut-off exponentially, but we note that including a realistic cut-off energy of tens of keV would only slightly move the pivot energy to lower values, as the small number of high energy photons do not strongly affect the pivot energy. The \citet{Uttley_2025} model uses linearized equations, which leads to a well-defined pivot energy as described by equation \ref{eq:pivotenergy}. However, as is clear from equation \ref{eq:pivotenergy}, changes in $\Gamma$, seed photon and heating luminosities and particularly seed photon energy $E_s$ will lead to different pivot energies. Because all these parameters vary slightly in time, the energy bands around the pivot energy are not linearly correlated with the other energy bands, leading to a decrease in coherence. In practice, seed photons will cover a range of energies, which could lead to variations in the average seed photon energy and thus the pivot energy. Above the pivot energy, the coherence recovers, but those energy bands will be anticorrelated with the softer bands.

However, in the framework of pivoting, it is very difficult to explain why the anticorrelation is only measured between very soft and very hard X-rays. A simple pivoting power-law would lead to anticorrelation above and below the pivot energy, i.e. on a broad range of energies, which is not observed.

\subsection{Hybrid spectral components}
\label{subsec:hybrid}

In this discussion, we have so far assumed that the coronal emission is well described by a single power-law with a high-energy cut-off. We know that this is an oversimplification with potentially significant consequences. For example, spectral-timing studies of MAXI J1820+070 by \citet{Yang_2025} revealed that the electron temperature associated with long-timescale variability is significantly higher than that on short timescales. The temperature on long timescales remains relatively constant throughout the hard state, whereas the short-timescale temperature evolves with X-ray luminosity. Two Comptonization components are suggested to account for those observed spectral-timing features. Also, in Figs. \ref{fig:lagcoh1} and \ref{fig:colormeshlog}, the low coherence for many combinations of energy bands may suggest that there are independently varying components with different contributions to the distinct energy ranges, leading to reduced coherence. 

As was shown by \citet{Zdziarski_2021hybrid} using \textit{{NuSTAR}} and \textit{INTEGRAL} data in the 3-650 keV energy range, the hard X-ray spectrum of \source{} cannot be explained by thermal Comptonization alone. Similar results were obtained for high-mass BHXRB Cyg X-1 \citep{McConnell_2002,Cangemi_2021}.
The presence of a hybrid plasma, also containing a power-law tail of non-thermal electrons, is inferred from the high energy spectrum above 80 keV. If the hard X-ray spectrum above $\sim$80 keV contains a large contribution of photons that are Compton-upscattered by a non-thermal electron distribution, the spectral-timing properties at these energies may be related to the nature of the non-thermal tail. \citet{Zdziarski_2021} generally assume a hot flow and truncated disc geometry for the corona, although their results do not depend on this assumption. An independent constraint on the hybrid plasma comes from the lack of an electron-positron annihilation feature around 511 keV, which means that the characteristic size of the corona must be $\gtrsim$4 $\rm{{R_g}}$. If we assume that the hybrid plasma is mainly located at the base of the jet, while a hot flow corona accounts for the power-law emission between $\sim2-30$ keV, complex interactions and imperfect signal propagation between the separate coronal components may lead to the complex pattern in the coherence we observe. 

A different explanation for the combination of a cut-off in the spectrum at around 100 keV and the power-law tail to higher energies is given by \citet{Beloborodov_2017}. In the model, the electrons in the corona are heated by two mechanisms: bulk motion of fast-moving plasmoids leads to up-scattering of photons up to hard X-ray energies of $\sim$100 keV, as observed. Additional harder X-rays are produced by electrons accelerated in magnetic reconnection events (known as X-points), which lead to the formation of the power-law tail \citep{Beloborodov_2017,Sironi_2020}. If those two mechanisms respond differently to e.g. seed photons from the disc, their behaviour may be related to the spectral-timing properties we observe in the following way.

If the non-thermal electron population creating the hard X-ray spectrum $\gtrsim80$ keV exchanges energy via a yet unknown mechanism with the accretion disc, which contributes to the spectrum $\lesssim1$ keV, it may lead to the observed anticorrelation. The dependence of the anticorrelated energies on the reference band may be due to the relative contributions of the disc and the low-energy end of the thermal-Comptonized emission, which overlaps with the disc blackbody energies. The soft thermal-Comptonized emission becomes relatively stronger towards higher energies until it dominates over the disc emission at $\gtrsim$1 keV. When the thermal-Comptonized contribution in the low-energy reference band is large, the anticorrelation of the disc with the non-thermal component will be suppressed at all energies where there is still significant thermal-Comptonized emission, such that one must go to higher hard X-ray energies to recover the anticorrelation. For this scenario to work, the seed photon population for the thermal corona cannot be the same as the one interacting with the non-thermal corona, which is in line with the reduced coherence we observe. In Fig. \ref{fig:stprops_0.3-0.5_3-10keV}, it is clear that the coherence is reduced at and below the QPO frequency between soft bands $\lesssim$1 keV and harder bands. If part of the disc variability is somehow filtered out by the QPO mechanism (see also \citealt{Bollemeijer_2025varlags}) and does not reach the thermal-Comptonized component, but still leads to anticorrelated behaviour in the non-thermal component, the observed energy-dependent lag and coherence may be reproduced. We note that the anticorrelation is only observed at low Fourier frequencies. At higher frequencies, the variability component that dominates medium energies between 1 and 30 keV (and does not show an anticorrelation) may also be dominant at very soft and very hard X-rays, effectively hiding the anticorrelated variability. However, we note that the large hard lags between the soft 0.3-0.5 keV reference band and the 100-250 keV band across a broad range of frequencies in Fig. \ref{fig:stprops_0.3-0.5_3-10keV} may also be related to the presence of both a correlated and an anticorrelated variability component at those frequencies. If both underlying signals contribute significantly at high energies, the result may be a low coherence and large lags from a combination of relatively small positive lags from the power-law and lags of $|\pi|$ rad from the anticorrelation, as is observed. The increase in the lags at higher energies and frequencies observed by \citet{Basak_2025} may be due to the same mechanism.

Finally, we mention that BHXRBs do not only emit in X-rays, but across the full electromagnetic spectrum. Timing studies with optical and infrared (OIR) and X-ray instruments allow studying the variability and measuring the lags and coherence between these different parts of the electromagnetic spectrum. In several sources, including \source, anticorrelations between OIR and X-ray variability are measured on time-scales of several seconds, while variability on shorter time-scales of tens of Hz is correlated \citep{Durant_2008,Paice_2019,Paice_2021,Vincentelli_2021}. The different types of (anti-)correlation on different time-scales are often attributed to different contributions from a hot flow and jet base corona. If the OIR emission arises due to synchrotron radiation that is also a source of seed photons for Compton-upscattering to hard X-rays, an anticorrelation between OIR and X-ray is produced naturally \citep{Veledina_2011}. A scenario in which soft X-rays below 1 keV are a source of seed photons for the hard emission above 100 keV, may lead to the anticorrelation we observe, possibly through the mechanisms investigated in Sections \ref{subsec:T_changes} and \ref{subsec:pivoting}. 

We als note that in reverberation lag studies in AGN, the degree of correlation (i.e. the coherence) between X-rays and ultraviolet (UV) or optical radiation has been measured. It was found that the correlation coefficient is lower than expected for a scenario in which X-rays from a static corona are reprocessed by the disc in and emitted in the UV or optical wavelengths \citep{Edelson_2015}. \citet{Panagiotou_2022_explain} found that a dynamic corona, of which the geometry varies on short timescales, can lead to the reduced X-ray vs UV/optical coherence observed. Reverberation frequencies in BHXRBs are thought to be much higher than the relatively long time-scales studied in this paper \citep{Uttley_2014review}. Still, the reduced coherence across a broad range of frequencies, especially for the hardest X-ray bands, may also be attributable to a time-dependent response. For example, if the geometry of the corona varies on short time-scales, as suggested by \citet{Bollemeijer_2024} and \citet{Bollemeijer_2025varlags}, the hard X-ray response may vary strongly as well, leading to the observed reduction in the coherence for higher energies in Fig. \ref{fig:stprops_0.3-0.5_3-10keV}. More detailed and complex models are required to test the validity of such an explanation.

\section{Conclusions}

Our findings can be summarised as follows:
\begin{enumerate}
    \item In BHXRB \source, we observe a phase lag of $|\pi|$ rad between the softest ($\lesssim$1 keV) and hardest ($\gtrsim$80 keV) X-rays on timescales of $\gtrsim$50 seconds. We interpret the phase lag as an anticorrelation between the soft reference band and the hardest X-ray bands.
    \item The anticorrelation is accompanied by a drop in coherence to very low values, which partially recovers at higher energies.
    \item The energy at which the phase lags go from 0 to $|\pi|$ rad and the coherence shows a drop depends linearly on the reference band energy (below 1 keV). Using harder reference bands, no anticorrelation is observed up to 250 keV, but the coherence decreases gradually above 30 keV.
    \item We consider three possible scenarios to explain our findings, but note that other, unexplored mechanisms may also play a role.
  \begin{enumerate}
    \item{Changes in the cut-off energy due to variations in e.g. seed photons can lead to an anticorrelation between soft and hard X-rays.  However, it is then unclear why the soft power-law emission from the corona, which is thought to respond to disc seed photon variations, does not show an anticorrelation with the hardest energy bands.}
    \item{A pivoting power-law can also lead to an anticorrelation at the observed energies. However, we would expect the anticorrelation to be visible between power-law energies above and below the pivot energy. In the data, only the very soft and hardest X-ray fluxes are anticorrelated, making this scenario unlikely.}
    \item{We speculate that a spectral component at hard X-ray energies, e.g. Comptonization by non-thermal electrons \citep{Zdziarski_2021}, is anticorrelated with the disc emission, exchanging energy with the disc via an unknown mechanism. If part of the low-frequency disc variability is filtered out by the QPO \citep{Bollemeijer_2025varlags} and is therefore not visible in the thermal Comptonized emission, this scenario would lead to the reduced coherence observed at low frequencies when using soft bands $\lesssim$1 keV and harder bands. Between $\sim$1 and 80 keV, coherent power-law variability dominates, while above and below those energies, the disc and non-thermal Comptonization components are important.} 
    \item{
    In any scenario, soft energy bands below 1 keV contain both disc emission and soft power-law photons and the relative strength of both types of variability depends on the exact energy bands used. The relative contribution from both spectral components may lead to the observed linear relation between the soft reference band and anticorrelated energies (the diagonal pattern in the lower right corners in Fig. \ref{fig:colormeshlog}). }
    \end{enumerate}
\end{enumerate}

\section*{Acknowledgements}
We thank the referee for their insightful comments and suggestions that have improved the paper. N.B. and this work are supported by the research program Athena with project No. 184.034.002, which is (partially) financed by the Dutch Research Council (NWO). B.Y. is supported by NSFC grants 12322307, 12361131579, 12273026; “the Fundamental Research Funds for the Central Universities”, Xiaomi Foundation / Xiaomi Young Talents Program.\\
This research makes use of the SciServer science platform (\url{www.sciserver.org}). SciServer is a collaborative research environment for large-scale data-driven science. It is being developed at, and administered by, the Institute for Data Intensive Engineering and Science at Johns Hopkins University. SciServer is funded by the National Science Foundation through the Data Infrastructure Building Blocks (DIBBs) program and others, as well as by the Alfred P. Sloan Foundation and the Gordon and Betty Moore Foundation. \\
This work made use of \textsc{astropy}:\footnote{\url{http://www.astropy.org}} a community-developed core Python package and an ecosystem of tools and resources for astronomy \citep{astropy:2013, astropy:2018, astropy:2022}. We also made use of \textsc{numpy} \citep{harris2020array}, \textsc{scipy} \citep{2020SciPy-NMeth} and \textsc{stingray} \citep{Huppenkothen2019a,Huppenkothen2019b,matteo_bachetti_2023_7970570}. 

\section*{Data Availability} This research has made use of data obtained through the High Energy Astrophysics Science Archive Research Center Online Service, provided by the NASA/Goddard Space Flight Center. The data underlying this article are available in HEASARC, at \url{https://heasarc.gsfc.nasa.gov/docs/archive.html}. 
This work has made use of data from the \hxmt{} mission, a project funded by China National Space Administration (CNSA) and the Chinese Academy of Sciences (CAS). All data are public and can be found at \url{http://archive.hxmt.cn/proposal}.
Upon publication, a basic reproduction package for the results and figures presented in this paper will be made available on Zenodo at \url{https://doi.org/10.5281/zenodo.16686671}.



\bibliographystyle{mnras}
\bibliography{ref_clean} 




\appendix

\section{List of simultaneous observations}
\label{app:simobs}

In table \ref{table:J1820data}, we show an overview of all simultaneous observations with \nicer{} and \hxmt{} during the bright hard state of the 2018 outburst. The reported simultaneous exposure times refer to the amount of time with good \nicer, ME and HE coverage. In the main body of the paper, we focus on epoch 2, which has by far the most simultaneous coverage. 

\begin{table*}
    \centering
    \begin{tabular}{c|c|c|c|c|c|c}
    \hline
        Epoch & ObsIDs NICER & ObsIDs HXMT & Date (2018) & MJD & Simultaneous time (ks) & QPO frequency (Hz) \\
        \hline\hline
        1 & 105 & 0201, 0202, 0203 & March 16 & 58193&  1.1  & - \\ 
        \hline
        2 & 107, 108, 109& 0301, 0302, 0304, 0307& March 22-27 & 58199-58204 & 18.2  & 0.04-0.05\\
        &110, 111, 112 &0401, 0402, 0403, 0404&&&\\
        &  & 0405, 0407, 0408, 0409&  & &  &  \\
        && 0410, 0415, 0416, 0601&&&\\
        \hline

        3 & 118, 120 & 1102, 1103, 1104, 1105 & April 1-4 & 58209-58212&6.0 & 0.06-0.07 \\
         & & 1201, 1202, 1203, 1204 &  & & &  \\
        \hline

        4 & 137 & 2801, 2802 & April 25 &58233& 3.1  & 0.17\\
        \hline

        5 & 156, 157 & 5301, 5302 & May 22-23 &58260-58261& 2.4  & 0.4-0.5 \\
         &  & 5401, 5402, 5403 & & & & \\
        \hline
        6 & 189 & 7801, 7802, 7803, 7804, 7805 & June 28 &58297& 6 & 0.42\\
        \hline
    \end{tabular}
    \caption{The six observation epochs with at least 1 ks of simultaneous data from \nicer{} and \hxmt{} ME and HE. The full \nicer{} ObsIDs are obtained by placing 1200120 before the three digits listed, while the full \hxmt{} ObsIDs can be found by adding P01146610.}
    \label{table:J1820data}
\end{table*}

\section{Errors on the coherence and the phase lags}
\label{sec:app_coherr}

The coherence function was first introduced well in astronomical time series literature by \citet{Vaughan_1997coherence}. The authors give clear prescriptions for the errors on the intrinsic coherence for three different situations: for high coherence and high data quality (small contribution from Poisson noise), low coherence and high data quality and low data quality. We find that for the included \nicer{} data, we can use the prescriptions for the first case. Since \citet{Vaughan_1997coherence} provide both relative and absolute errors, which may easily lead to mistakes, we show the formulae for the coherence and the absolute errors on the coherence that we use in the paper.

We use for the raw coherence $\gamma_{\rm{r}}^2$

\begin{equation}
    \gamma_{\rm{r}}^2 =\frac{|C_{12} |^2}{P_1 P_2 },
\end{equation}

where $C_{12} $ is the cross-spectrum of energy bands 1 and 2 averaged over $K$ frequencies and $M$ light curve segments, while $P_1 $ and $P_2 $ are the average power spectra of those energy bands, without noise subtraction. The absolute errors on the raw coherence are given by

\begin{equation}
    \Delta\gamma_{\rm{r}}^2 = \sqrt{2/(KM)}(1-\gamma_{\rm{r}}^2 )|\gamma_{\rm{r}} |.
\end{equation}

If we also have a good estimate of the Poisson noise level, we can calculate the intrinsic coherence $\gamma_{\rm{i}}^2$ using

\begin{equation}
    \gamma_{\rm{i}}^2 =\frac{|C_{12} |^2-b^2}{(P_1 -n_1)(P_2 -n_2)},
\end{equation}
where $n_1$ and $n_2$ are the Poisson noise levels in the power spectra and $b^2$ is a bias term \citep{Vaughan_1997coherence}, defined as

 \begin{equation}
     b^2=\left[(P_1 -n_1)n_2 + (P_2 -n_2)n_1 +n_1 n_2 \right] /(KM).
 \end{equation}. 

 Because estimation of the bias correction itself is noisy, it is recommended to include it only when $KM<500$ \citep{Ingram_2019error}. We note that the type of normalization of the power spectrum is not important for the calculation of the coherence, as long as it is the same in all power- and cross-spectra used. The absolute errors on the intrinsic coherence then become

\begin{equation}\label{eq:intcoherr}
   \begin{multlined}
     \Delta\gamma_{\rm{i}}^2  = \frac{\gamma_{\rm{i}}^2 }{\sqrt{KM}}\\
     \Bigg[ 
     \frac{2b^2KM}{(|C_{12} |^2-b^2)^2} 
     + \frac{n_1^2}{(P_1 -n_1)^2}  \\
     + \frac{n_2^2}{(P_2 -n_2)^2}
     + \frac{KM(\Delta\gamma^2_{\rm{r}} )^2}{(\gamma^2_{\rm{i}} )^2   } 
     \Bigg] ^{1/2},
  \end{multlined}
\end{equation}

For lag- and coherence versus energy spectra, we follow the approach described by \citet{Ingram_2019error} in their equation 17. The expressions above are valid for high signal-to-noise data with medium to high values of coherence, as is the case for most of the data we show in this paper. For lower signal-to-noise, where the source variability is less than a few percent of the Poisson noise level, expression \ref{eq:intcoherr} generally underestimates the errors. Because no better prescription is currently known, we still use the above equations, but keep in mind that the errors can be underestimated.
\newline

We calculated the errors on the phase lags for the lag-energy spectra in Fig. \ref{fig:lagcoh1} in two ways. The thin black error bars are calculated using equation 19 in \citet{Ingram_2019error}, which take into account the fact that variability in different energy bands is correlated. Still, however, the errors on the phase lags are much larger than the scatter between individual data points, especially above 2 keV. The overestimation arises from the fact that the (raw and intrinsic) coherence between the 0.3-0.4 keV reference band and the harder energy bands, which is used to calculate the errors, is low ($\sim0.25$). The energy bands between 2 and 30 keV are strongly correlated, as their coherence is close to unity (see lower right panel of Fig. \ref{fig:lagcoh1}), so the scatter on the measured lags is much smaller than predicted when assuming a coherence of 0.25 between the individual harder bands as well, which is an implicit assumption in \citet{Ingram_2019error}. 

To set a lower limit on the error bars, we also calculated them while assuming unity coherence between all bands. By following equations 43 and 46 in \citet{Ingram_2019error} and assuming intrinsic coherence $\gamma_i^2=1$, we find the following lower limit on the error bars:
\begin{equation}
    \mathrm{d}\tilde{\phi}(s)=\sqrt{\frac{\tilde{P}_{\mathrm{r}} P_{\mathrm{s,noise}}}{2KM|\tilde{C}(s)|^2}}.
\end{equation}

\section{Telemetry gaps in \nicer{} data}
\label{app:gaps}
\begin{figure}
    \centering
    \includegraphics[width=\linewidth]{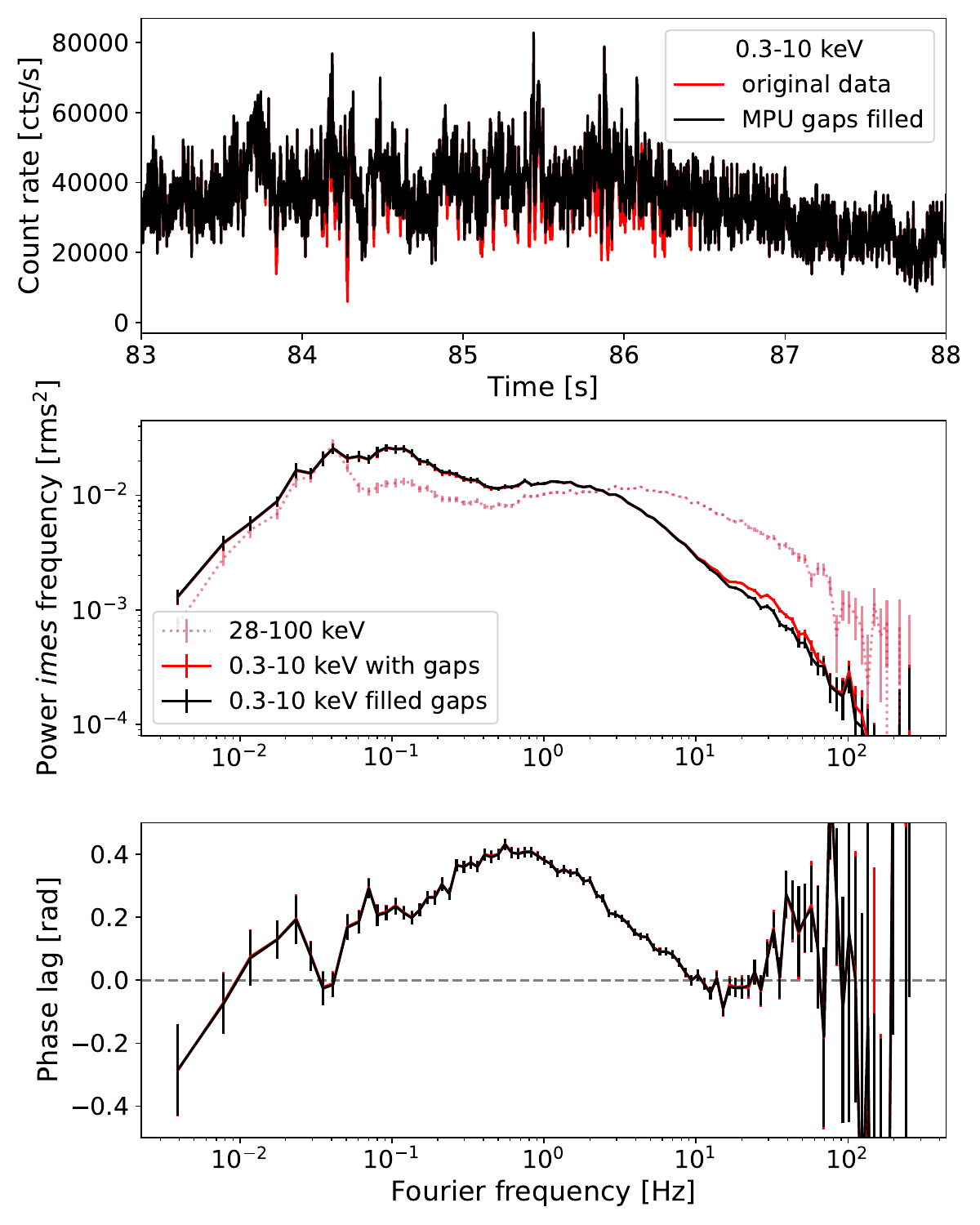}
    \caption{The upper panel shows an example of a 0.3-10 keV light curve in red, with clear spurious variability due to telemetry saturation. Because MPUs switch on and off independently, the total count rate does not reach 0, but light curves of individual MPUs do. The black light curve shows the effect of filling short gaps in each MPU light curve and effectively removes the extra variability at high count rates. Both light curves have a time resolution of $\frac{1}{512}$ s. The lower panels show the very limited influence of the gaps on power spectra and lags for all data in epoch 2. The power spectrum is affected above $\sim$20 Hz, while the lags show no visible difference. The lower frequencies addressed in this work are not affected.}
    \label{fig:fillgaps}
\end{figure}

The high source brightness and low absorption of \source{} led to telemetry saturation within \nicer. \nicer{} consists of 52 functioning Focal Plane Modules (FPMs) grouped into seven Measurement and Power Units (MPUs). Telemetry saturation occurs at the MPU level, causing MPUs to stop registering events for a fraction of a second, leading to fragmented GTIs and short gaps in the light curve.\footnote{\url{https://heasarc.gsfc.nasa.gov/docs/nicer/data_analysis/nicer_analysis_tips.html}} Such spurious variability affects Fourier properties and could in principle lead to anticorrelations, as a high source brightness leads to artificial drops in the measured count rates. To test whether these issues affect our conclusions, we created light curves per MPU and filled all gaps between GTIs shorter than 1 s with values drawn from a Poisson distribution with $\lambda=\langle x_{\rm{short}} \rangle$, with $\langle x_{\rm{short}} \rangle$ defined as the mean count rate in GTIs shorter than 1 s (i.e. the highest count rates in a given observation, which lead to fragmented GTIs). The effect of filling the short gaps in light curves from individual MPUs can be seen in Fig. \ref{fig:fillgaps}. Correcting for the telemetry saturation this way removes spurious variability at high frequencies ($\sim20$ Hz), but does not visibly affect Fourier properties at low frequencies. We conclude that the low-frequency anticorrelation is not caused by data artefacts. 

Some of the individual FPMs are affected to such an extent that they are filtered out by \texttt{nicerl2}. We ensured that within each observation, the number of included FPMs stayed constant, to avoid spurious swings in count rate not related to the source. The number of FPMs used ranges from 14 to 28 out of a maximum of 52. In our analysis, we normalised all light curves to what would have been observed with 52 functioning FPMs before calculating Fourier products. Although this increases the Poisson noise contribution in observations with smaller numbers of detectors, not correcting for the number of FPMs leads to a spurious drop in the coherence when calculating cross-spectral properties for \nicer{} and \hxmt{} data. 

\bsp	
\label{lastpage}
\end{document}